# Diffraction-free space-time beams

H. Esat Kondakci and Ayman F. Abouraddy[*]

*CREOL, The College of Optics & Photonics, University of Central Florida, Orlando, FL 32816, USA*

*corresponding author: raddy@creol.ucf.edu

**Diffraction-free optical beams propagate freely without change in shape and scale. Monochromatic beams that avoid diffractive spreading require two-dimensional transverse profiles, and there are no corresponding solutions for profiles restricted to one transverse dimension. Here, we demonstrate that the temporal degree of freedom can be exploited to efficiently synthesize one-dimensional pulsed optical sheets that propagate self-similarly in free space. By introducing programmable conical (hyperbolic, parabolic, or elliptical) spectral correlations between the beam's spatio-temporal degrees of freedom, a continuum of families of axially invariant pulsed localized beams is generated. The spectral loci of such beams are the reduced-dimensionality trajectories at the intersection of the light-cone with spatio-temporal spectral planes. Far from being exceptional, self-similar axial propagation is a *generic* feature of fields whose spatial and temporal degrees of freedom are tightly correlated. These one-dimensional 'space-time' beams can be useful in optical sheet microscopy, nonlinear spectroscopy, and non-contact measurements.**

Diffractive spreading is a fundamental feature of freely propagating optical beams that is readily observed in everyday life. Diffraction sets limits on the optical resolution in microscopy, lithography, and photography; on the maximum distance for free-space optical communications and standoff detection; and on the precision of spectral analysis[1,2]. As a result, there has been a long-standing fascination with so-called 'diffraction-free' beams whose change in shape and scale during propagation is curbed when compared to other beams of comparable transverse size[3]. Monochromatic diffraction-free beams have sculpted two-dimensional (2D) transverse spatial profiles that confirm to Bessel[4], Mathieu[5], or Weber[6] functions, among other examples (see Refs. [7,8] for recent taxonomies). The situation is altogether different for monochromatic beams with *one* transverse dimension – or optical sheets – where there are *only two* possible diffraction-free solutions: the cosine wave that lacks spatial localization and the Airy beam that maintains a localized intensity profile but whose center-of-mass undergoes a transverse shift with propagation[9,10]. Indeed, a conclusive argument by Michael Berry[11] identified the Airy beam as the *only* such monochromatic one-dimensional (1D) profile. Optical nonlinearities can be exploited to thwart diffractive spreading[12,13], and in some cases chromatic dispersion is required in the medium to restrain the diffraction of pulsed beams[14-16]. However, most applications require *free-space* diffraction-free beams.

Here, we exploit the temporal degree of freedom (DoF) *in conjunction* with the spatial DoF to realize a variety of diffraction-free pulsed solutions having arbitrary 1D transverse profiles. By establishing a correlation between the spatial and temporal DoFs, diffractive spreading is reined-in and the time-averaged spatial profile propagates self-similarly. We call such pulsed-beam solutions 'space-time' (ST) beams since the salient characteristic that enables this behavior is the tight correlation between the spatial and temporal DoFs. Previous studies of pulsed 2D diffraction-free beams (or 'localized waves'[3]) have explored the cases of separable spatio-temporal DoFs (a Bessel transverse profile and an Airy-shaped pulse, for example[17]) and those of correlated DoFs, including Brittingham's focus wave modes[18], X-waves[19-22], Bessel pulses[23], and Mackinnon's non-dispersive wave packets[24]. Such polychromatic fields are superpositions of 2D diffraction-free Bessel beams at each wavelength[3,7]. Thus, *no* experimental demonstrations of 1D diffraction-free beams *in free space* have been reported to date.

The 1D diffraction-free ST-beams we synthesize here may be viewed from a unified perspective in the ST spectral domain, where we take the spatial frequencies (or transverse wave numbers) $k_x$ and the temporal frequencies $\omega$ as the independent variables. Programmable correlations can be efficiently introduced between $k_x$ and $\omega$ via phase-only modulation in an optical configuration comprising a diffraction grating and a spatial light modulator that assigns each spatial frequency to a different wavelength. In this conception, the spectral loci of ST-beams are reduced-dimensionality trajectories at the intersection of the light-cone with surfaces in the ST spectral domain, such as the conic sections at the intersection of the light-cone with tilted spectral *planes*. One such plane is the iso-plane $\mathcal{P}_0$ of the axial wave number, whose intersection with the light-cone is a branch of a hyperbola that corresponds to beams with no axial ST-dynamics[25-27]. Planes tilted with respect to $\mathcal{P}_0$ intersect with the light-cone in hyperbolae, parabolas, or ellipses – all of which are spectral loci for families of diffraction-free ST-beams. This ST spectral-tunability, moreover, helps modify the proportionality between the temporal and spatial spectral widths required to synthesize ST-beams.

Using this strategy, we create a 1D pulsed beam (monitored in real space and in the ST spectral domain) with a transverse width of the time-averaged intensity of 7 µm that propagates for 2.5 cm, and another of width 14 µm that propagates for 10 cm. Furthermore, by modulating the ST spectral-phase, we also produce a diffraction-free pulsed beam in the form of a 'hollow sheet' having a field minimum at its center. Such 1D diffraction-free ST-beams can be useful in optical sheet microscopy, nonlinear and multi-modality spectroscopy, standoff detection of chemicals, and high-resolution bar-code readers.

**Theory for diffraction-free ST-beams**

The theoretical underpinning for 1D diffraction-free ST-beams can be elucidated by reference to the $(k_x, k_z, \omega)$ spectral-space in Fig. 1, where monochromatic plane waves $e^{i(k_x x + k_z z - \omega t)}$ are represented by points on the surface of the light-cone $k_x^2 + k_z^2 = \omega^2/c^2$. We use positive frequencies $\omega$ in one half of a double-napped light-cone, and take only the positive-$k_z$ spatial spectrum, corresponding to forward (causal) propagation. The ST spectra we consider are confined to conic sections at the intersection of the light-cone with various spectral planes. For example, the spatial spectra of monochromatic beams lie on circles at the intersection of the light-cone with iso-frequency planes $\omega = \omega_0$ (Fig. 1a), whose plane-wave expansion is $E(x, z; t) = e^{-i\omega_0 t} \int dk_x \tilde{E}(k_x) e^{i(k_x x + k_z z)}$, such that all the spatial frequencies are assigned to the same frequency (Fig. 1b). Here, diffractive spreading is induced by axial dephasing along $z$ due to the different $k_z$ associated with each $k_x$ (Fig. 1c). A *pulsed beam* corresponds in general to a 'patch' on the light-cone (Fig. 1d) whose plane-wave expansion is

$$E(x, z; t) = \iint dk_x d\omega \tilde{E}(k_x, \omega) e^{i(k_x x + k_z z - \omega t)}, \tag{1}$$

such that each $k_x$ is assigned to multiple frequencies (Fig. 1e); $\tilde{E}(k_x, \omega)$ is the Fourier transform of $E(x, 0; t)$. In general, such a beam diffracts and the associated pulse disperses (Fig. 1f). It is typical to consider *separable* ST-spectra $\tilde{E}(k_x, \omega) = \tilde{E}_x(k_x)\tilde{E}_t(\omega)$, in which case the field also separates in space and time $E(x, 0; t) = E_x(x)E_t(t)$. Nevertheless, $x$ and $t$ naturally couple to each other upon free propagation and $E(x, z; t)$ is thus no longer separable[1,28]. This so-called 'space-time coupling' is particularly substantial in the case of focusing of ultra-short pulses, and pre-conditioning the ST-spectrum $\tilde{E}(k_x, \omega)$ is required to diminish the undesirable impact of such coupling[29-31].

Key to generating diffraction-free ST-beams in our conception is exploiting *1D ST-spectral trajectories* on the light-cone, which have reduced dimensionality in comparison to the *2D patch* in Fig. 1d. The distinguishing feature of ST spectra confined to such trajectories is that $|k_x|$ and $\omega$ have a one-to-one correlation. Considering first a propagation-invariant ST-solution of the form $E(x, z; t) = E(x, 0; t)e^{i\beta z}$, the 1D wave equation imposes a *hyperbolic* dispersion constraint,

$$\beta^2 = \omega^2/c^2 - k_x^2, \tag{2}$$

where $\beta$ is a constant axial wave number[26,27]. This correlation function between $|k_x|$ and $\omega$ results from intersecting the light-cone with an iso-$k_z$ plane $\mathcal{P}_0$ ($k_z = \beta$) in a single branch of a hyperbola (Fig. 1g). The plane-wave expansion of a beam whose spectrum is confined to this hyperbola is[26,27]

$$E(x, z; t) = e^{i\beta z} \int dk_x \tilde{E}(k_x) e^{i(k_x x - \omega(k_x)t)}, \tag{3}$$

in which case the propagation dynamics of the ST-beam is *independent* of $z$ except for an overall phase. The ST-spectrum $\tilde{E}(k_x, \omega) = \tilde{E}(k_x)\delta(\omega - \omega(k_x))$ assigns each $k_x$ to a unique $\omega$ (Fig. 1h), and the ST-intensity profile in the time-frame of the pulse is independent of $z$ (Fig. 1i). The idealized delta-function ST spectral-correlation renders the beam energy, diffraction length, and group velocity formally infinite. However, these quantities all become finite once the delta-function correlation is relaxed and replaced with a sharply peaked function centered at $\omega = \omega(k_x)$ with finite width $\delta\omega$. This apodization results from the inevitable spectral uncertainty induced in the correlation function as a consequence of restricted aperture sizes, finite number and size of pixels in modulators, and limited time-detection windows[32].

In addition to the hyperbola in Fig. 1g, other conical sections feature self-similar propagation characteristics. Tilting the spectral plane an angle $\varphi$ in the $(k_z, \omega/c)$-plane with respect to the $(\omega/c)$-axis – while maintaining the vertex at $(k_x, k_z, \omega/c) = (0, \beta, \beta)$ – produces intersections with the light-cone in the form of hyperbolae $\beta^2 = \omega^2/c^2 - \eta k_x^2$ in the $(k_x, \omega/c)$-plane while $\varphi < 45°$, where $\eta = \eta(\varphi)$ leads to changes in the curvature of the hyperbola. Further increasing $\varphi$ leads to intersections in the form of parabolas ($\varphi = 45°$) or ellipses ($\varphi > 45°$), as in Fig. 1j, where the sign of the curvature may change (Fig. 1k). We find that ST-beams associated with this variety of trajectories are indeed all pseudo-diffraction-free (Fig. 1l).

**Experimental setup**

We now address the challenge of efficiently synthesizing optical beams with the above-described ST-correlations. The basic concept – as illustrated in Fig. 2 – combines spatial-beam modulation[33,34] and ultrafast pulse-shaping[35,36], and is related to the so-called $4f$-imager utilized to introduce ST-coupling into ultrafast pulsed beams for nonlinear spectroscopy and quantum control[37-39]. We start from a pulsed plane-wave $E(x, 0; t) = E(t)$, whose separable ST-spectrum is $\tilde{E}(k_x, \lambda) \approx \tilde{E}(\lambda)\delta(k_x)$, and spread the spectrum along the $y$-axis with a diffraction grating, so that each wavelength $\lambda$ is assigned to a position $y(\lambda)$. In lieu of a 1D phase-modulation of the spectrum only along $y$ to reshape the pulse[35,36], we exploit the orthogonal spatial dimension $x$ in conjunction with $y$ and implement a 2D phase modulation $\exp\{i\psi(x,y)\}$. At location $y(\lambda)$ a linearly varying phase along $x$ is implemented whose slope corresponds to a specific $k_x$. By controllably associating each $k_x$ with a unique $\lambda$, programmable correlations are introduced into the ST-spectrum $\tilde{E}(k_x, \lambda)$. Recombining the spectrum via a second grating reconstitutes the pulse and *simultaneously* overlaps all the spatial frequencies, so the ST-beam is immediately realized along $x$ with no need for a spatial Fourier transform[40,41].

A generic pulsed beam (100-fs FWHM, ≈10-nm bandwidth centered at a wavelength of 800 nm) is dispersed by a grating and directed to a 2D spatial light modulator (SLM) that imprints a phase of the form $\psi(x, y) = (|x - x_c|/x_s)(y/\alpha y_s)^{1/2}$ mod $2\pi$; where $\alpha$ defines the slope of the hyperbolic – or other conic – trajectories, $x_s$ and $y_s$ are scaling factors depending on the SLM details (pixel size, phase calibration, and wavelength spread across the SLM), and $x_c$ is the central point on the SLM along the vertical direction. Each row on the SLM imprints a spatial frequency corresponding to the wavelength incident at that row. The correlation between $|k_x|$ and $\lambda$ is one-to-one except for the ambiguity across the width of the individual SLM pixels, which contributes to a spectral uncertainty $\delta\lambda$ in this correlation. The modulated beam is reflected back and the pulse is reconstituted by the grating to produce the ST-beam. To characterize the beam we employ a two-lens telescope system that transports the ST-beam into one of two measurement modalities: characterization in physical space $(x, z)$ by a CCD camera translated axially along $z$ to capture the time-averaged 1D intensity profile $I(x, z) \propto \int dt |E(x, z; t)|^2$ at different $z$-planes; and characterization in the ST spectral domain $|\tilde{E}(k_x, \lambda)|^2$ via a second CCD after performing a spatial Fourier transform operation along $x$ and inserting a grating to spread the spectrum along $y$ (jointly performed with the help of a spherical lens).

**ST-beams in the spectral domain**

We first confirm that we obtain the target ST parametric relationships $\omega = \omega(k_x)$ corresponding to the various conical sections at the intersection of the light-cone with ST spectral planes. We plot in Fig. 3 the measured ST spectral intensities $|\tilde{E}(k_x, \lambda)|^2$ for four distinct beams. The first (Fig. 3a-c) corresponds to a randomized parametric relationship that eliminates correlations between the spatial and temporal DoFs. A random SLM phase pattern (Fig. 3b) produces a separable ST-spectrum $|\tilde{E}(k_x, \lambda)|^2 \sim |\tilde{E}_x(k_x)|^2 |\tilde{E}_t(\lambda)|^2$ (Fig. 3c). The spatial frequencies in the random SLM phase pattern are selected independently for each wavelength from a Gaussian probability distribution whose FWHM is 0.2 rad/μm. The spatial bandwidth $\Delta k_x = 0.2$ rad/μm generates a beam of transverse spatial width $\Delta x = 16.4$ μm. Such a beam (corresponding to the patch in Fig. 1d) undergoes diffraction upon free propagation.

The spectrum of the second ST-beam lies on the iso-$k_z$ hyperbolic trajectory at the intersection of the light-cone with the plane $\mathcal{P}_0$, $k_z = \beta$ (Fig. 3d), which enforces a one-to-one correlation between $|k_x|$ and $\lambda$, with the larger spatial frequencies assigned to higher temporal frequencies (shorter wavelengths); Fig. 3e. The measured spectral intensity in Fig. 3f fits well the theoretical curve $\beta^2 = \omega^2/c^2 - k_x^2$, with $\Delta k_x = 0.55$ rad/μm (corresponding to $\Delta x = 7.6$ μm) and $\Delta \lambda = 0.32$ nm. ST-beams lying on this spectral locus have $\Delta \lambda$ correlated to $\Delta k_x$, and a narrower beam profile thus necessitates a shorter accompanying pulse. However, this proportionality can be altered *without* impacting the diffraction-free characteristics of the ST-beam by tilting the plane $\mathcal{P}_0$.

The third ST-beam (Fig. 3g-i) is associated with a hyperbolic spectral trajectory of different curvature at the intersection of the light-cone with a spectral plane $\mathcal{P}_\varphi$ tilted an angle $\varphi$ with respect to $\mathcal{P}_0$ while remaining parallel to the $k_x$-axis and passing through the point $(k_x, k_z, \omega/c) = (0, \beta, \beta)$; where $\beta = \omega_o/c$ and $\omega_o$ is the maximum or minimum frequency in the spectrum (the vertex of the conic section); Fig. 3g. The equation of $\mathcal{P}_\varphi$ is $k_z = \beta + (\omega/c - \beta)\tan\varphi$, and its intersection with the light-cone is a hyperbola when $\varphi < 45°$. The synthesized ST-beam in Fig. 3i corresponds to $\varphi \approx 30°$, such that the resulting hyperbola is $\beta^2 = \omega^2/c^2 - 4k_x^2$, which has a higher curvature when compared to the iso-$k_z$ beam in Fig. 3f. This is clear in the SLM phase pattern (Fig. 3h), where the larger temporal bandwidth $\Delta\lambda = 0.5$ nm is associated with a smaller spatial bandwidth $\Delta k_x = 0.28$ rad/μm. Tuning $\varphi$ thus helps adjust the temporal bandwidth associated with the spatial bandwidth. By further increasing the tilt angle $\varphi$, the hyperbola on the light-cone becomes a parabola at $\varphi = 45°$, and then an ellipse when $\varphi > 45°$. The fourth synthesized ST-beam in Fig. 3j-l corresponds to a section of an ellipse on the light-cone at $\varphi \approx 60°$ (Fig. 3j). The sign of the curvature of this trajectory has been reversed with respect to those in Fig. 3f and Fig. 3i, such that larger $k_x$ is assigned to lower $\omega$, as is clear in the reversed phase pattern in Fig. 3k along the $\lambda$-axis (see also Fig. 1k). The measured $|\tilde{E}(k_x, \lambda)|^2$ fits well the equation of the ellipse $\beta^2 = \omega^2/c^2 + 4k_x^2$.

**ST-beams in physical space**

Our experimental arrangement is thus capable of continuously tuning across the full gamut of conic-sectional trajectories that are spectral loci for diffraction-free ST-beams. We now proceed to monitoring ST-beam-propagation in *physical* space to confirm diffraction-free behavior. We find that diffraction upon free propagation is considerably curtailed once the DoFs are correlated. The width of the transverse profile remains inversely proportional to $\Delta k_x$, but the effective Rayleigh range $z_R$ is substantially increased. Several examples are shown in Fig. 4. A first example in Fig. 4a shows a beam whose measured time-averaged profile $I(x, z)$ has a width of $\Delta x \approx 14$ μm propagating for $\approx 100$ mm (10 cm) along the $z$ axis, corresponding to $\approx 140\times$ the $z_R$ of a Gaussian beam of FWHM 14 μm, after which the axial intensity drops sharply. The locus of the ST spectrum for this beam is the hyperbolic trajectory in Fig. 3i (the intersection of the spectral plane $\mathcal{P}_{30}$ with the light-cone), whereupon $\Delta k_x = 0.28$ rad/μm and $\Delta\lambda = 0.5$ nm. By employing a beam whose ST-spectrum is confined instead to the elliptical trajectory in Fig. 3l having the same curvature (albeit opposite sign), we obtain the same result in Fig. 4a.

Further reducing the transverse width of the beam requires increasing $\Delta k_x$, which in turn necessitates the use of a larger spectrum $\Delta\lambda$. However, varying the tilt angle $\varphi$ of the spectral plane $\mathcal{P}_\varphi$ can help change the proportionality between $\Delta k_x$ and $\Delta\lambda$. We verify this in the measurements presented in Fig. 4b where the width of the beam profile is reduced to $\Delta x \approx 7$μm by increasing the spatial bandwidth to $\Delta k_x = 0.55$ rad/μm. Nevertheless, by exploiting the iso-$k_z$ plane $\mathcal{P}_0$ instead of $\mathcal{P}_{30}$, this larger $\Delta k_x$ is associated with a *smaller* temporal bandwidth of $\Delta\lambda = 0.32$ nm. The resulting ST-beam propagates for $\approx 25$ mm before the axial intensity drops sharply, corresponding to $\approx 140\times$ the $z_R$ of a Gaussian beam of FWHM 7 μm. The required $\Delta\lambda$ can be further reduced for a beam with the same or even smaller $\Delta x$ by tilting $\mathcal{P}_0$ by negative $\varphi$. Indeed, measurements of multiple ST-beams whose ST spectra are confined to the hyperbolic trajectories in the planes $\mathcal{P}_0$ and $\mathcal{P}_{30}$ show that although the dependence of $\Delta\lambda$ on $\Delta k_x$ has changed (Fig. 4d), nevertheless, the inverse dependence of $\Delta x$ on $\Delta k_x$ remains the same for ST-beams located in either plane (Fig. 4e), and – in principle – for ST-beams lying in all other spectral planes. In general, the ratio of the spectral-correlation uncertainty $\delta\lambda$ to the bandwidth $\Delta\lambda$ (that is, $\Delta\lambda/\delta\lambda$) sets the diffraction-free length[26], which is in general orders-of-magnitude larger than that of a Gaussian beam of the same transverse width.

The reliance of the arrested diffractive-spreading on the ST correlation is confirmed in Fig. 4c where a beam having uncorrelated DoFs (Fig. 3c) displays normal diffraction behavior. The DoFs are uncoupled, and a beam of minimum width $\Delta x \approx 16.4$ μm results from the spatial bandwidth $\Delta k_x \approx 0.2$ rad/μm, corresponding to $z_R \approx 750$ μm.

**Diffraction-free space-time hollow sheet**

It is important to note that the 1D conic-section trajectories on the light-cone (Fig. 3d,g,i) represent only the *loci* of pulsed optical beams. By assigning a complex field *amplitude* to each point on the light-cone, an infinity of pulsed beams can be synthesized associated with each trajectory (even while holding $\Delta k_x$, $\Delta \lambda$, and $\delta \lambda$ fixed). In our arrangement, this can be achieved by modulating the field magnitude and phase at each wavelength (or spatial frequency, since $k_x$ and $\lambda$ are tightly correlated) before or after incidence on the SLM (Fig. 2b). Following our theme of phase-only modulation, we present an example in Fig. 5 of a hollow optical sheet (or a pair of closely placed sheets) realized by modulating the beam spectrum with a phase factor of the form $\exp\{i\frac{\pi}{2}u(k_x)\}$, where $u(\cdot)$ is the unit-step function; that is, a $\pi$-phase-step is introduced between the two halves of the spatial spectrum around $k_x = 0$, which thereby becomes an odd function (Fig. 5a,b). The spectral locus of this ST-beam lies on the hyperbolic iso-$k_z$ trajectory in Fig. 3f. The measured time-averaged intensity $I(x,z)$ shows the expected null at the origin propagating axially with arrested diffraction for ≈ 25 mm while maintaining a width of ≈ 8.5 μm; Fig. 5c.

Varying the phase-step between the two halves of the spectrum gradually transitions the ST-beam from that in Fig. 4b (where the beam center is a peak) to that in Fig. 5c (where the beam center is a null). We plot in Fig. 5d,e the profile $I(x,z)$ at fixed $z$ while varying $\theta$ in a phase factor of the form $\exp\{i\frac{\theta}{2}u(k_x)\}$ imparted to the ST spectrum of the two beams shown in Fig. 4b,c with $\Delta x = 7$ μm and 14 μm, respectively. Such a phase pattern smoothly tunes the 'parity' of the spectrum[42]. Notably, while the ST-beam profile becomes spatially asymmetric around $x = 0$ as $\theta$ is tuned (Fig. 5d,e), the beam nevertheless propagates along $z$ with arrested diffractive spreading and *no* transverse shift.

## Conclusion

We have shown that introducing spatio-temporal correlations into a pulsed beam – controllably and efficiently implemented with a phase-only modulation scheme – produces diffraction-free beams having one transverse dimension. This scheme can be readily extended to beams with two transverse dimensions by replacing $k_x$ with $\left(k_x^2 + k_y^2\right)^{1/2}$, in effect 'rolling-up' the extra dimension along the light-cone and replacing each 1D cosine wave with a 2D Bessel wave. The trajectories of the diffraction-free spectral loci we have explored here are all conical sections at the intersection of the light-cone with tilted spectral planes. This approach does not exhaust all the possibilities. Because the one-to-one correlation between $|k_x|$ and $\omega$ is the essence of the diffraction-free intensity of the time-averaged beam, unexplored trajectories generated by a variety of connected surfaces intersecting with the light-cone can also yield such beams. It remains an open question which of these trajectories preserve the ST-profile in the pulse time-frame *besides* maintaining the time-averaged intensity profile. It appears at the moment that only those trajectories having *linear* projections in the $(k_z, \omega)$-plane exhibit this property. This approach full control of the beam profile at the pulse center, but more work is needed to exercise similar control over the spatial profile of the time-averaged profile. The diffraction-free beams described here provide opportunities for profoundly new applications in imaging[43], nonlinear optics[44], and plasma and filamentation studies[45]. Furthermore, the underlying principle of ST-beams is equally applicable to other physical waves, such as acoustic, electron, and matter waves, and potentially to elementary particle physics[46].

Spatio-temporal correlations in optics manifest themselves in a variety of guises: the space-time coupling common to tight-focusing of ultra-short pulses[29,30] and the generation of ultra-intense laser pulses[47], among a host of other examples. In most cases, this ST coupling is unintentional, requiring careful characterization and pre-conditioning of the beam to obviate its impact[31]. Here we have exploited such correlations to arrest diffractive spreading of ST-beams in free space. Indeed, axially self-similar propagation becomes a *generic* feature of beams incorporating these correlations in their ST-spectrum. This work can thus be viewed as a generalization of burgeoning working on so-called 'classical entanglement'[48-50] to continuous DoFs of the optical field. Previous demonstrations of classical entanglement have been confined to date to discretized DoFs such as polarization and individual spatial or temporal modes. We have extended this concept to continuous spatio-temporal DoFs and shown that it underlies the observed diffraction-free propagation.

**Acknowledgments**. We thank D. N. Christodoulides and R. Menon for helpful discussions. This work was supported by the U.S. Office of Naval Research (ONR) under contract N00014-14-1-0260.

# Figures and captions

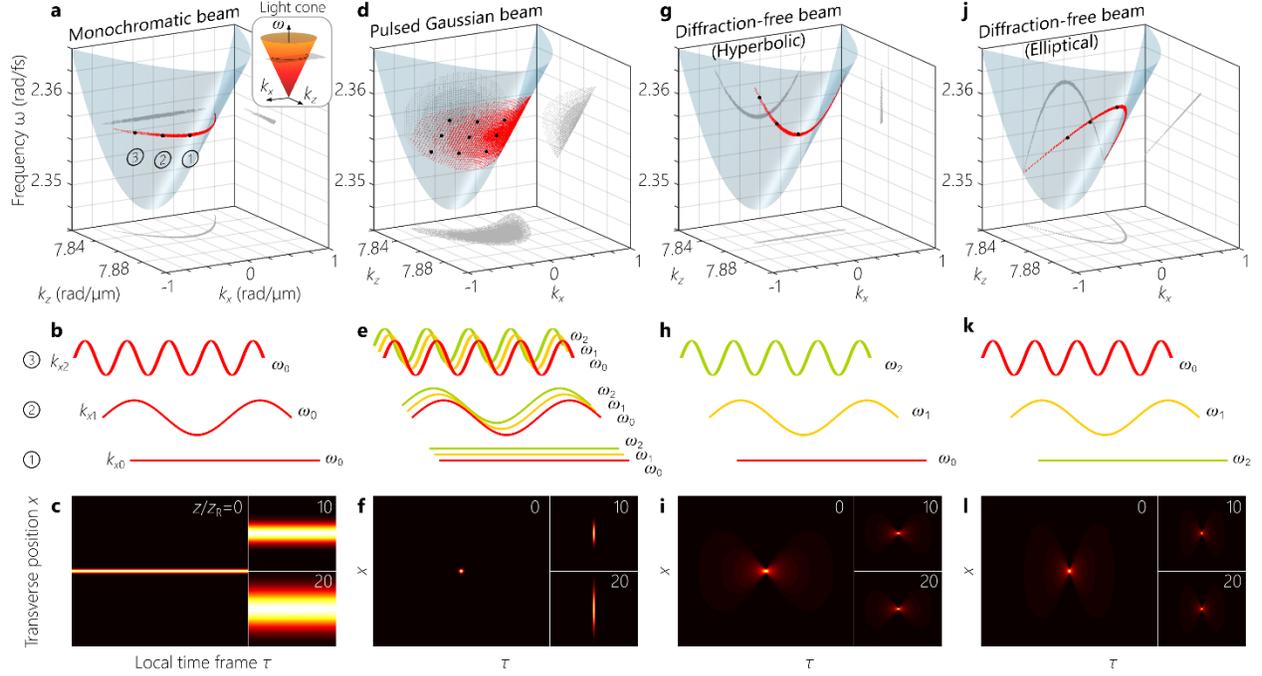

**Figure 1 | Concept of a diffraction-free space-time beam.** Representing 1D beams in $(k_x, k_z, \omega)$-space: each point on the light-cone surface $k_x^2 + k_z^2 = \omega^2/c^2$ corresponds to monochromatic plane wave. **a**, Monochromatic beams lie on the intersection of the light-cone with a horizontal iso-frequency plane (inset shows the coordinate system). The projections in the $(k_x, \omega)$-plane and $(k_z, \omega)$-plane is each a line $\omega = \omega_0$ and in the $(k_x, k_z)$-plane an arc from a circle $k_x^2 + k_z^2 = \omega_0^2/c^2$. **b**, The spatial frequencies ($k_{x0}$, $k_{x1}$, and $k_{x2}$), corresponding to the black dots on the light-cone surface in (**a**), are all associated with the same frequency $\omega_0$. **c**, Calculated ST-intensity profiles $|E(x, z; t)|^2$ in the local time-frame $\tau$ of the beam in the planes $z = 0$, $10z_R$, and $20z_R$, showing diffractive spreading with free propagation; $z_R$ is the Rayleigh range. **d**, A pulsed Gaussian beam corresponds to a patch on the light-cone. **e**, Each spatial frequency is now associated with many frequencies. **f**, Calculated $|E(x, z; t)|^2$ at $z = 0$, $10z_R$, and $20z_R$ showing diffractive spreading. **g**, A spectral locus for ST-beams at the intersection of the light-cone with the vertical iso-$k_z$ plane. The projection in the $(k_z, \omega)$-plane is a line $k_z = \beta$ and in the $(k_x, \omega)$-plane a branch of a hyperbola $\omega^2/c^2 - k_x^2 = \beta^2$. **h**, Each spatial frequency is associated with a unique frequency (large $k_x$ associated with high $\omega$). **i**, Calculated $|E(x, z; t)|^2$ at $z = 0$, $10z_R$, and $20z_R$ showing propagation-invariance with $z$. **j**, A spectral locus for ST-beams at the intersection of the light-cone with a plane parallel to the $k_x$-axis and tilted an angle $\varphi$ with respect to the $\omega$-axis. The projection in the $(k_z, \omega)$ plane is a line with slope $\tan\varphi$ and in the $(k_x, \omega)$-plane a section of an ellipse $\omega^2/c^2 + \eta k_x^2 = \beta^2$. **k**, Each spatial frequency is associated with a unique frequency (small $k_x$ associated with high $\omega$). **l**, Calculated $|E(x, z; t)|^2$ at $z = 0$, $10z_R$, and $20z_R$ showing propagation-invariance with $z$. Throughout we have $\lambda_0 = 800$ nm, $\Delta\lambda = 1$ nm, $\delta\lambda = 0.05$ nm, and $\Delta x = 7$ μm at $z = 0$ and $\tau = 0$.

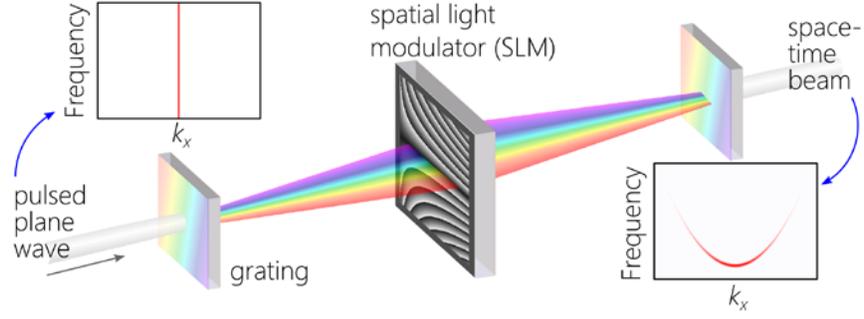

**Figure 2 | Synthesis and analysis of diffraction-free space-time beams.** Concept of ST-beam synthesis. The spectrum of a pulsed plane-wave is spread by a diffraction grating before impinging on a 2D spatial light modulator (SLM) that modulates the wave front with a 2D phase distribution $\psi(x, y)$. Each wavelength $\lambda$ is imbued with a spatial frequency $k_x(\lambda)$. The insets on the left and right show the ST-spectra $|\tilde{E}(k_x, \omega)|^2$ before and after ST-beam synthesis.

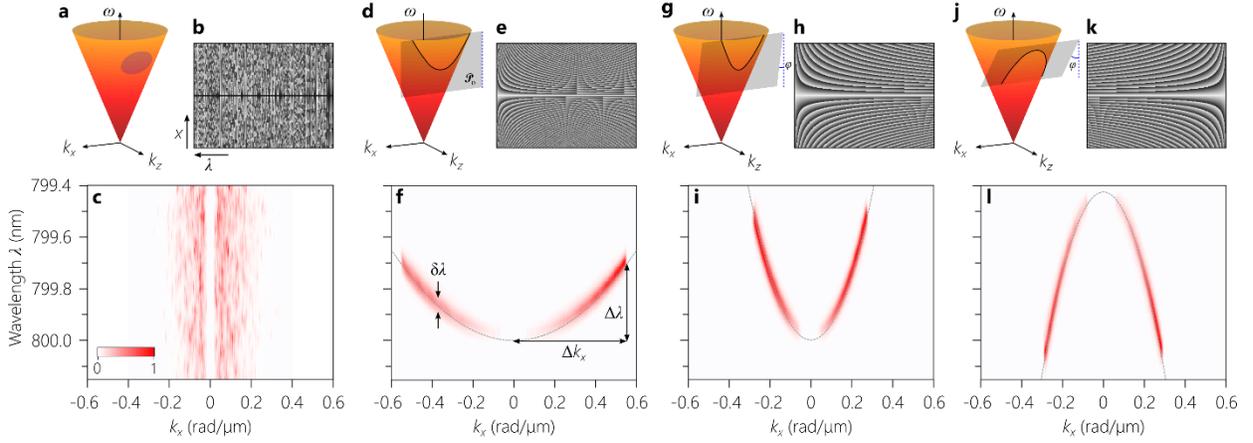

**Figure 3 | Spatio-temporal spectral control of ST-beams. a-c**, A beam with randomized (uncorrelated) DoFs. **a**, Depiction of the beam on the light-cone. **b**, SLM phase pattern used to produce the ST-spectrum. **c**, The measured spectral intensity $\left|\tilde{E}(k_x, \lambda)\right|^2$. **d-f**, ST-beam with *hyperbolic* spectral correlation whose locus is at the intersection of the light-cone with an iso-$k_z$ plane $\mathcal{P}_0$. **f**, The dotted black line is a guide for the eye and corresponds to a single branch of the hyperbola $\beta^2 = (\omega/c)^2 - k_x^2$; the bandwidth is $\Delta\lambda = 0.32$ nm, the correlation uncertainty has a FWHM of $\delta\lambda = 50$ pm, and the spatial bandwidth is $\Delta k_x = 0.55$. **g-i**, ST-beam with *hyperbolic* spectral correlation at the intersection of the light-cone with a tilted plane $\varphi \approx 30°$, $\Delta\lambda = 0.5$ nm, $\delta\lambda = 70$ pm, and $\Delta k_x = 0.28$ rad/μm. The dotted black line corresponds to a single branch of the hyperbola $\beta^2 = (\omega/c)^2 - 4k_x^2$. **j-l**, ST-beam with *elliptical* spectral correlation at the intersection of the light-cone with a tilted plane $\varphi \approx 60°$, $\Delta\lambda = 0.62$ nm, $\delta\lambda = 70$ pm, and $\Delta k_x = 0.3$ rad/μm. The dotted black line corresponds to a section of the ellipse $\beta^2 = (\omega/c)^2 + 4k_x^2$.

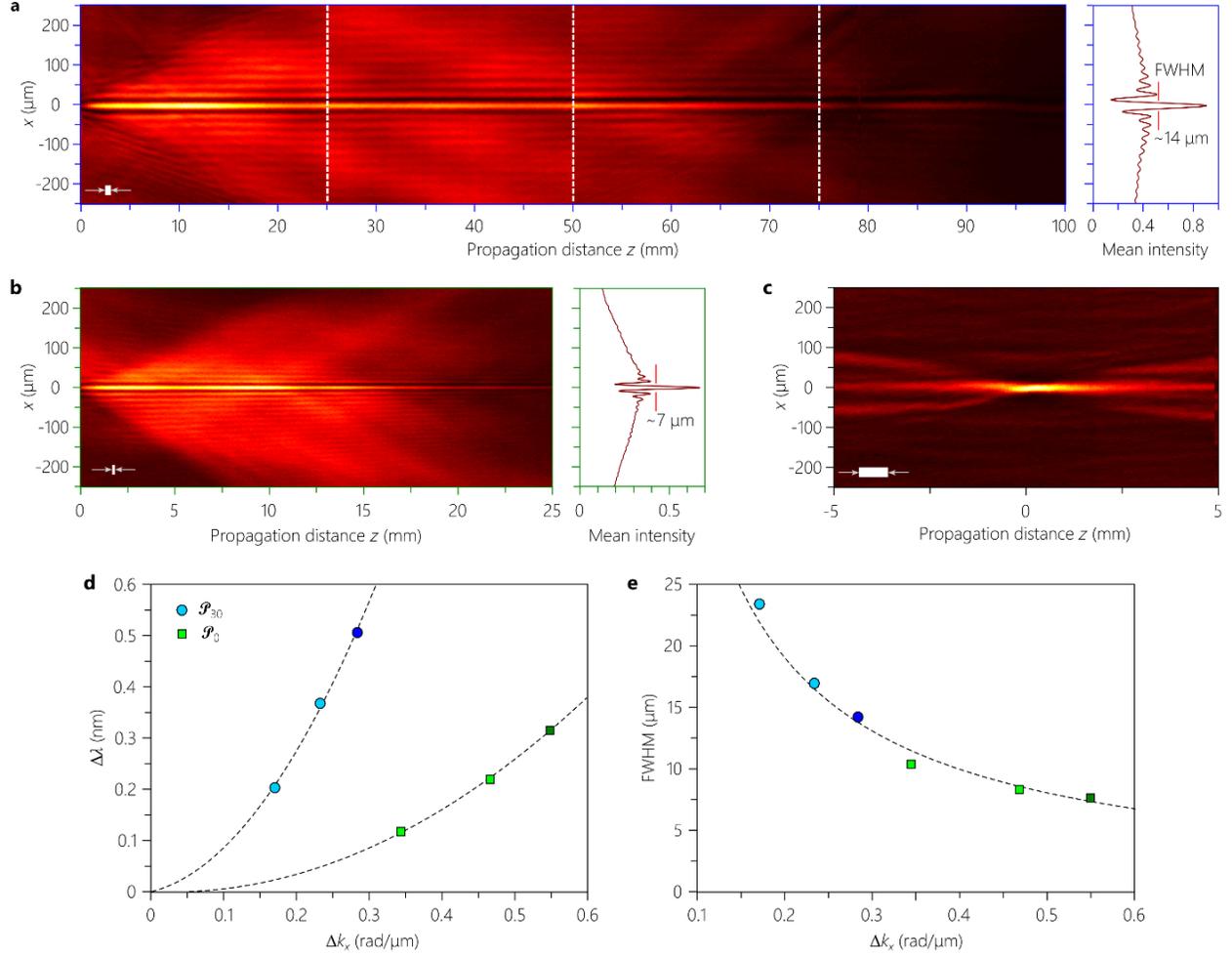

**Figure 4 | Propagation of diffraction-free space-time beams. a**, Measured ST-beam time-averaged intensity in real space $I(x,z)$ corresponding to the ST-spectrum in Fig. 3i. The width of the central feature in the transverse profile is ≈14 μm (FWHM) and is preserved for ≈ 10 cm. The solid white line in the bottom-left corner is the Rayleigh range $z_R = 555$ μm for a Gaussian beam with a FWHM of 14 μm. The vertical dashed white lines marks the stitching of multiple data sets for successive 25-mm-long axial scans. The right panel depicts the axially averaged transverse intensity profile $\int I(x,z)dz$ over the 10-cm propagation distance. The tail of the spatial profile exhibits a slow decay. **b**, Measured $I(x,z)$ for the ST-spectrum in Fig. 3f, which produces a 7-μm-FWHM central spatial feature preserved for ≈ 25 mm. The thin white line in the bottom-left corner is the Rayleigh range $z_R = 136$ μm for a Gaussian beam with a FWHM of 7 μm. The decay of the tail in $\int I(x,z)dz$ is faster, which implies a more optimal design of the ST-spectrum. **c**, Measured $I(x,z)$ for the uncoupled ST-spectrum in Fig. 3c, which produces a 16.4-μm-FWHM central spatial feature preserved for the standard Rayleigh range $z_R = 750$ μm of a Gaussian beam with the same FWHM (the white segment in the bottom-left corner). **d**, Measured relationship between the temporal bandwidth $\Delta\lambda$ and the spatial bandwidth $\Delta k_x$ for ST-beams synthesized according to the hyperbolic spectral trajectories in Fig. 3f (green squares, spectral plane $\mathcal{P}_0$) and Fig. 3i (blue circles, spectral plane $\mathcal{P}_{30}$). The circle and square having darker colors correspond to the ST-beams in (**a**) and (**b**), respectively. The dashed lines are parabolic fits. **e**, Measured dependence of the FWHM of the ST-beam on the spatial bandwidth $\Delta k_x$. The green squares and blue circles correspond to ST-beams produced by the ST-spectra in Fig. 3f and Fig. 3i, respectively. The dashed line is a fitted $1/x$ relationship.

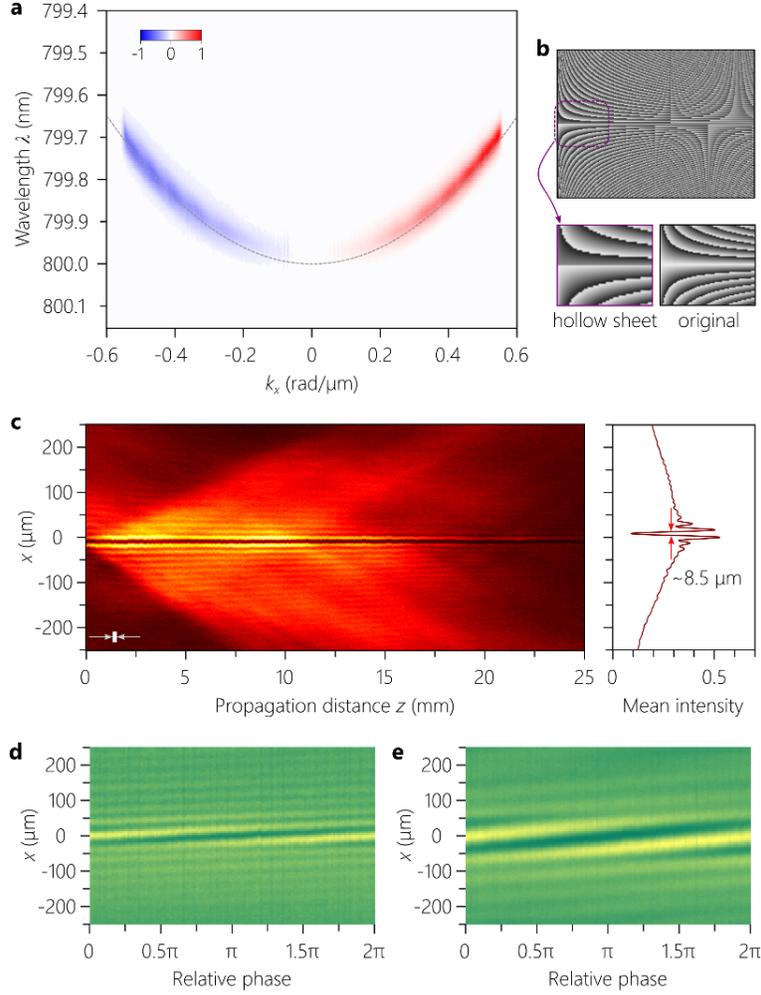

**Figure 5 | Diffraction-free space-time hollow-sheet beams. a**, False-color image of the measured ST spectral intensity $|\tilde{E}(k_x,\omega)|^2$ for a 1D hollow sheet highlighting the $\pi$-phase-step implemented along $k_x$. **b**, The SLM phase pattern used to create the ST-spectrum in (**a**). The left inset draws attention to the impact of the $\pi$-phase-step required to implement the hollow-sheet in (**c**) in comparison to the right inset associated with the ST-beam in Fig. 4b. **c**, Measured $I(x,z)$ for the 1D hollow-sheet beam. The FWHM of the dip is 8.5 μm, which is preserved for an axial distance of $\approx 25$ mm. The right panel depicts the axially averaged transverse intensity profile $\int I(x,z)dz$ over the 25-mm propagation distance. **d-e**, Transverse intensity of a ST-beam $I(x,0)$ after modulating the ST-spectrum along $k_x$ with a phase-factor of the form $\exp\{i\frac{\theta}{2}u(k_x)\}$, where $u(\cdot)$ is the unit step-function. The center of the ST-beam $I(x=0)$ gradually transitions while varying the relative phase $\theta$ from a peak ($\theta = 0$; Fig. 4a-c) to a dip ($\theta = \pi$; Fig. 5a-c). **d**, The spectral locus of the ST-beam lies on the hyperbolic trajectory in Fig. 3f. **e**, The spectral locus of the ST-beam lies on the hyperbolic trajectory in Fig. 3i.